\documentclass[aps,twocolumn,groupedaddress,showpacs]{revtex4-1}
\usepackage{graphicx}

\begin{document}

\title{Condensate wave function and elementary excitations of bosonic polar
molecules: beyond the first Born approximation}

\author{Chao-Chun Huang$^{1}$, Daw-Wei Wang$^{2,3}$, and Wen-Chin Wu$^1$}

\affiliation{$^1$ Department of Physics, National Taiwan Normal
University, Taipei 11650, Taiwan
\\
$^2$ Physics Department, National Tsing-Hua University, Hsinchu 300,
Taiwan
\\
$^3$ Physics Division, National Center for Theoretical Sciences,
Hsinchu 300, Taiwan}

\begin{abstract}
We investigate the condensate wave function and elementary
excitations of strongly interacting bosonic polar molecules in a
harmonic trap, treating the scattering amplitude beyond the
standard first Born approximation (FBA). By using an appropriate
trial wave function in the variational method, effects of the
leading order correction beyond the FBA have been investigated and
shown to be significantly enhanced when the system is close to the
phase boundary of collapse. How such leading order
effect of going beyond the FBA can be observed in a realistic
experiment is also discussed.
\end{abstract}
\maketitle
\section{Introduction}

In recent years dipolar gases become a fast growing field of
theoretical and experimental interests in the studies of ultracold
atoms and molecules. Among several dipolar systems, chromium
($^{52}$Cr) atoms have been successfully realized and studied to
great extents
\cite{griesmaier05,stuhler05,fattori06,griesmaier06,lahaye07,koch08}.
The dipolar interaction effects for $^{87}$Rb atoms as well as
$^{39}$K atoms are also observed in different groups
\cite{Rb_Berkly,K_Florence}. Several polar molecule systems such
as CO \cite{doyle99}, ND$_3$ \cite{bethlem00}, RbCs \cite{sage05},
LiCs \cite{deiglmayr08}, and CsCl have also been trapped, cooled,
and studied \cite{doyle04,egorov04}. More recently, a high
phase-space density gas of polar $^{40}$K$^{87}$Rb molecules have
also been produced \cite{ni08}. These stimulate great interest in
the studies of dipolar systems at low temperatures. In an earlier
theoretical work within the first Born approximation (FBA), Yi and
You \cite{yi00} proposed a pseudopotential to study the long-range
dipolar interaction. Based on this approximation, which is
justified only in the weak dipole moment limit, various
theoretical studies of the excitations, collapses, instability,
{\em etc}. of the dipolar BEC system have been carried out in
these years \cite{goral02}.

While Yi and You's pseudopotential within FBA is appropriate for a
weakly interacting system, it can not be applied for a polar
molecule system of strong interaction. Under a strong field, polar
molecules can have a large electric dipole moment to make various
shape resonances possible. Therefore one needs to include the
interaction beyond the FBA in order to have a better understanding
on the low energy behavior of these dipolar systems. Recently, one
of us has developed an effective many-body theory for bosonic
polar molecules in the strong interaction regime that goes beyond
the FBA \cite{wang08}. It is interesting to investigate how the
higher order interactions affect the properties of polar
molecules. Following \cite{wang08}, our current paper attempts to
study the leading order effects beyond the FBA on the condensate
profile and elementary excitations of bosonic polar molecules. It
will be shown that the effect of the leading-order term is most
prominent when the system is approaching the instability
(collapsed state). This is manifested in both the ground-state
property and elementary excitations. In particular, at the
intermediate trap aspect ratio, $\lambda
=\omega_z/\omega_\rho=5\sim 6$, the results obtained by the
theories within FBA and going beyond FBA have a drastic difference
near the collapsed regime. This drastic change indeed allows one
to extract quantitatively the effect beyond FBA in a realistic
experiment. In the large $\lambda$ (pancake) and small $\lambda$
(cigar) limits, on the other hand, the effects beyond the FBA
becomes much reduced and may not be easily observed in the
experiment.

The paper is organized as follows. In Sec.~II, we outline the
effective Hamiltonian for studying the low-energy many-body
physics of polar molecules.  In Sec.~III,
a trial wave function, called ``modified Gaussian" is introduced
and used to calculate the ground-state properties of polar
molecules beyond the FBA. In Sec.~IV, we investigate the
elementary excitations (breathing modes) of the system, and provide
an scheme to quantitatively extract the effect beyond the FBA in an
experiment. Sec.~V is a brief conclusion.

\section{Effective Hamiltonian}\label{sec2}
\label{sec21}

The effective Hamiltonian, which describes low-energy many-body
physics of bosonic polar molecules beyond the FBA, can be given in
the following second quantization formalism (for details, see
Ref.~\cite{wang08}):

\begin{eqnarray}
H_{\rm eff}&=&\sum\limits_{\bf{p}} {\left( {{\varepsilon
_{\bf{p}}} - \mu } \right) \hat a_{\bf{p}}^\dag {{\hat
a}_{\bf{p}}}}  + \frac{1}{\Omega
}\sum\limits_{{{\bf{p}}_{\bf{1}}},{{\bf{p}}_{\bf{2}}}} {\hat
a_{{{\bf{p}}_{\bf{1}}}}^\dag {{\hat
a}_{{{\bf{p}}_{\bf{2}}}}}{V_{\rm ext}}
\left({\bf p}_1-{\bf p}_2\right)} \nonumber  \\
&+&\frac{1}{{2\Omega }}\sum\limits_{{{\bf{p}}_{\bf{1}}},
{{\bf{p}}_{\bf{2}}},{\bf{P}}} {\hat a_{\frac{1}{2}{\bf{P}} +
{{\bf{p}}_{\bf{1}}}}^\dag \hat a_{\frac{1}{2}{\bf{P}} -
{{\bf{p}}_{\bf{1}}}}^\dag {{\hat a}_{\frac{1}{2}{\bf{P}} -
{{\bf{p}}_{\bf{2}}}}}{{\hat a}_{\frac{1}{2}{\bf{P}} +
{{\bf{p}}_{\bf{2}}}}}\Gamma
\left({\bf p}_1,{\bf p}_2\right)}, \nonumber \\
\label{eq:Heff}
\end{eqnarray}
where $V_{\rm ext}({\bf p}_1-{\bf p}_2)$ is the Fourier transform of
the external trapping potential and $\Omega$ is the system volume.
$\Gamma( {\bf p}_1,{\bf p}_2)$ is
the pseudo-potential responsible for the interaction vortex
between polar molecules, and can be
divided into the following three parts:
\begin{eqnarray}
\Gamma({\bf{p}},{\bf{p'}})= \frac{4\pi \hbar^2
a_s}{M}+V_d({\bf{p}}-{\bf{p'}})-\frac{4\pi \hbar^2}{M} f_\Delta
({\bf{p}},{\bf{p'}}). \label{eq:Gamma2}
\end{eqnarray}
The first term in Eq.~(\ref{eq:Gamma2}) is from the standard
$s$-wave scattering and the second term is due to the usual FBA
with $V_d({\bf{q}})={4\pi {D^2}}({3{{\cos }^2}{\theta_{q_z}} -
1})/3$ being the long-range dipole-dipole interaction. Here $D$ is
the electric dipole moment and $\theta_{q_z}$ is the angle between
the $z$-component wavevector and the total wavevector. The third
term, $f_\Delta$, is the scattering amplitude including all the
results deviated from the known FBA in the second term. In the
general situation, the deviated term can be expanded in different
angular momentum channels~\cite{wang08},
\begin{eqnarray} f_\Delta({\bf p}_1,{\bf
p}_2)\equiv -4\pi\sum_{l,l^\prime}i^{l^\prime -l}\sum_m
\Delta{a}_{ll^\prime}^{(m)}Y_{lm}^\ast(\hat{p}_1) Y_{l^\prime
m}(\hat{p}_2), \label{eq:f-Delta}
\end{eqnarray}
where $\Delta{a}_{ll^\prime}^{(m)}\equiv -i^{l^\prime
-l}[t_{lm}^{l^\prime m}(0)-t_{Blm}^{l^\prime m}(0)]$ is the
difference between a full scattering length and its FBA result.
$Y_{lm}(\hat{p})$ is the spherical harmonics defined by the
orientation of momentum ${\bf p}$. In Eq.~(\ref{eq:f-Delta}), the
sum, $\sum_{l,l'}$, has excluded the contribution from a pure
$s$-wave channel ($l=l'=0$), which has been already included in
the first term of Eq.~(\ref{eq:Gamma2}). In the limit of a weak
external field, the electric dipole moment is also small and each
term in Eq.~(\ref{eq:Gamma2}) has the following orders of
magnitude: $a_s ={\cal O}(D^0)$, $V_d ={\cal O}(D^2)$, and
$\Delta{a}_{ll^\prime}^{(m)}={\cal O}(D^4)$ \cite{wang08}. Thus in
the low field ({\em i.e.}, small-$D$) limit, $f_\Delta
({\bf{p}},{\bf{p'}})$ term can be safely neglected within the FBA.
However, when the external field is strong enough, the effect of
$f_\Delta ({\bf{p}},{\bf{p'}})$ should be taken into account to go
beyond the standard FBA. The actual values of $\Delta
a_{ll'}^{(m)}$ have to be calculated from a full scattering theory
of polar molecules, as shown in Ref.~\cite{deb01,Bohn,Blume}.

In the present paper, we shall study how these higher order terms,
$\Delta{a}_{ll^\prime}^{(m)}$, can affect the ground state and the
elementary excitation property of a bosonic dipolar molecule gas
beyond the first Born approximation level. According to the
numerical calculation of these scattering amplitude in
Refs.~\cite{deb01,Blume}, we find that $\Delta a_{ll'}^{(m)}$
becomes smaller for larger angular momentum, $l$ or $l'$. When near
the first shape resonance, it is found that although all scattering
channels will diverge, but the most dominant contributions are still
from $(l,l')=(0,0)$ ($s$-wave), and $(0,2)=(2,0)$ channels.
Divergences on the $(2,2)$ channels are almost invisible. When away
from the shape resonance regime, the scattering amplitude of higher
angular momentum channels ($l>0$) are well described by the first
Born approximation, if only the dipole moment is not too large.
Therefore, from a practical point of view, we do not need to
investigate all the possible values of $\Delta a_{l,l'}^{(m)}$, but,
on the other hand, we can concentrate on the effect of the lowest
non-trivial terms, $\Delta{a}_{0,2}^{(0)}\neq 0$ and set
$\Delta{a}_{ll^\prime}^{(0)}=0$ for all $(l,l^\prime)$ not equal to
$(0,2)$, $(2,0)$, or $(0,0)$.

As mentioned above, this special case
becomes relevant when the electric dipole moment is close to the
first shape resonance peak, where the coupling between $s$-wave and
$d$-wave channels are significantly enhanced. The effect of shape
resonance on the $s$-wave channel, $a_s$, is also significant, but
has been separated in Eq.~(\ref{eq:Gamma2}). When the dipole moment
is even larger, it is reasonable to expect that more scattering
channels will have scattering amplitude different from the first
Born approximation result, i.e. $\Delta a_{ll'}^{(m)}\neq 0$.
However, since it is very difficult to investigate a general
properties of such strong interacting limit, here we will still
concentrate on the effects beyond the FBA only through the channel,
$\Delta a_{0,2}^{(0)}\neq 0$, which will be a good approximation
when the external field is near the regime of the first shape
resonance \cite{deb01}. Nevertheless, the effect of next leading
term, $\Delta{a}_{2,2}^{(0)}$, will be briefly studied in Sec.~IV.

To calculate the expectation value of the effective Hamiltonian
Eq.~(\ref{eq:Heff}), $E=\langle H_{\rm eff}\rangle$, one can
replace $\hat a_{\bf{k}}$ by a macroscopic condensate
wave function, $\Psi_{\bf{k}} \equiv \langle \hat a_{\bf{k}}
\rangle= \frac{1}{\sqrt{\Omega}}\int
d{\bf{r}}\Psi({\bf{r}})\,e^{-i{\bf{k}}\cdot{\bf{r}}}$ in
(\ref{eq:Heff}) at zero temperature. We can then apply the
variational method to obtain the ground-state energy of the
system. If one uses the (normalized) simple Gaussian-type trial
wave function ($\rho^2=x^2+y^2$),
\begin{eqnarray}
\Psi(\bf r)&=&\frac{\exp(-\rho^2/2R_0^2-z^2/2R_z^2)}
{\pi^{3/4}R_0R_z^{1/2}}, \label{eq:trialsg}
\end{eqnarray}
with $R_0$ and $R_z$ the Gaussian radii of the condensate in the
$x$-$y$ plane and along the $z$ axis respectively and assumes that
the harmonic trapping potential is $V_{\rm ext}({\bf
r})=\frac{1}{2}m\omega_\rho^2 \rho^2 +\frac{1}{2}m\omega_z^2z^2$
with $\omega_\rho$ and $\omega_z$ the trapping frequencies,
variational energy of Eq.~(\ref{eq:Heff}) becomes (see also
Eq.~(16) of \cite{wang08})
\begin{eqnarray}
\frac{{E({R_0},{R_z})}}{{{E_0}}} =E_{\rm k}+E_{\rm trap}+E_{\rm
int},
\label{eq:simpleg21}
\end{eqnarray}
where
\begin{eqnarray}
E_{\rm k} = \frac{{R_0^2 + 2R_z^2}}{{4R_z^2R_0^2}},
\label{eq:Ek}
\end{eqnarray}
\begin{eqnarray}
E_{\rm trap}= \frac{{2R_0^2 + {\lambda
^2}R_z^2}}{4},
\label{eq:Etrap}
\end{eqnarray}
and
\begin{eqnarray}
E_{\rm int}= \frac{N}{{\sqrt {2\pi } R_0^3}}\left[
{{a_s}\frac{{{R_0}}}{{{R_z}}} + 8\left( {\frac{{{a_d}}}{{3\sqrt 5
}} - \Delta a_{0,2}^{(0)}} \right){A_2}(\frac{{{R_0}}}{{{R_z}}})}
\right]\label{eq:Eint}
\end{eqnarray}
correspond to the kinetic energy, trapping potential energy, and
interaction energy respectively. Note that the above results include
only the leading higher-order term, $\Delta a_{0,2}^{(0)}$. Here
$A_l(\frac{R_0}{R_z})\equiv\frac{\sqrt{2l+1}}{8} \int_{-1}^1 dx
\frac{P_l(x)}{(1+((R_0/R_z)^2-1)x^2)^{3/2}}$ and
$\lambda\equiv\omega_z/\omega_\rho$ is the trapping aspect ratio.
All lengths, $a_s$, $a_d\equiv MD^2/\hbar^2$, $\Delta
a_{0,2}^{(0)}$, $R_0$, and $R_z$, are scaled by the harmonic
oscillator length, $a_{\rm osc,0}\equiv \sqrt{\hbar/M\omega_\rho}$.
$E_0\equiv N\hbar^2/m a_{\rm osc,0}^2$ is the energy scale.
Throughout this paper, we will use the same length and energy
scales.

It is convenient to redefine: ${\zeta _s} \equiv N{a_s}, {\zeta
_d} \equiv N{a_d}$, and ${\zeta _{0,2}}\equiv -3\sqrt 5  N{\Delta
a_{0,2}^{(0)}}$ such that the interaction term Eq.~(\ref{eq:Eint})
can be rewritten as
\begin{eqnarray}
E_{\rm int}= \frac{1}{{\sqrt {2\pi } R_0^3}}\left[
{{\zeta_s}\frac{{{R_0}}}{{{R_z}}} + {8\over 3\sqrt{5}}(\zeta_d +
\zeta_{0,2}){A_2}(\frac{{{R_0}}}{{{R_z}}})}
\right].\label{eq:Eint2}
\end{eqnarray}
In view of Eq.~(\ref{eq:Eint2}) or (\ref{eq:Eint}), it poses a
subtlety that both the dipolar interaction ${\zeta _d}$ and the
higher-order interaction ${\zeta _{0,2}}$ couple to the same
function. This result is just an artifact of the Gaussian trial
wave function and should not exist in a more general condensate
wave function as stated in Ref.~\cite{wang08}.

\section{Beyond the FBA: ground-state property}\label{sec3}

In order to separate the contribution of the $\Delta a_{0,2}^{(0)}$ for the
usual FBA result, in this paper we use the following
``modified Gaussian" (MG) trial wave function
\begin{eqnarray}
 \Psi ({\bf r}) &= &C \exp(-\rho^2/2R_0^2
 -z^2/2R_z^2)\nonumber\\
  &\times& \left( {1 + \frac{{a_0 \rho^2 }}{{R_0^2 }} +
 \frac{{a_z z^2 }}{{R_z^2 }}}\right), \label{eq:wave}
 \end{eqnarray}
where $R_0$, $R_z$, $a_0$, and $a_z$ are variational parameters and
$C=2\pi ^{-3/4}/[R_0 \sqrt{R_z}( 8a_0^2 + 4{a_0}{a_z} + 3a_z^2 +
8{a_0} + 4{a_z} + 4)]$ is the normalization factor. Function
(\ref{eq:wave}) can be viewed as the simple Gaussian (SG) multiplied
(modified) by a parabolic function. In Ref.~\cite{huang07}, a
similar MG trial function (SG function multiplied
by a hypercosine function) was first introduced and used
to study one- and two-component ultracold BEC systems.
When the MG trial wave function, Eq.~(\ref{eq:wave}), is used to
calculate the energy functional, $E=E(R_0,R_z,a_0,a_z)$,
ground-state energy of the system can be obtained by minimizing $E$.
That is, it requires that $\partial E/\partial \beta_i=0$ are
satisfied for all four variational parameter
$\beta_i=R_0,R_z,a_0,a_z$. Comparison between the results due to a
SG and a MG trial wave function for one- and two-component ultracold
BEC systems without a dipolar
interaction can be found in Ref.~\cite{huang07}.

To justify the appropriateness of MG trial wave function (\ref{eq:wave})
in the variational studies of strong dipolar BEC systems,
we refer to the following.
Recent numerical studies on dipolar BEC systems
have revealed structured ground-state density profiles in them
\cite{ronen07,dutta07}. Of most interest, density
profile can exhibit a double-peak structure, so-called
``biconcave condensates", for dipolar BECs.
In a recent work of Jiang and Su \cite{jiang06}, it
has been shown that the structured double-peak ground state density profile
can be successfully reproduced in the variational approach based on a
similar trial wave function, analogous to (\ref{eq:wave}).
Thus it is believed that, at least in the qualitative manner,
MG trial wave function should be a good one
for studying the higher-order effect beyond the FBA in
strongly interacting dipolar BEC systems. It is hoped that
a more accurate numerical approach, which is difficult but is
underway, can be completed soon.

\begin{figure}
\vspace{0.1cm}
\includegraphics[width=0.5\textwidth]{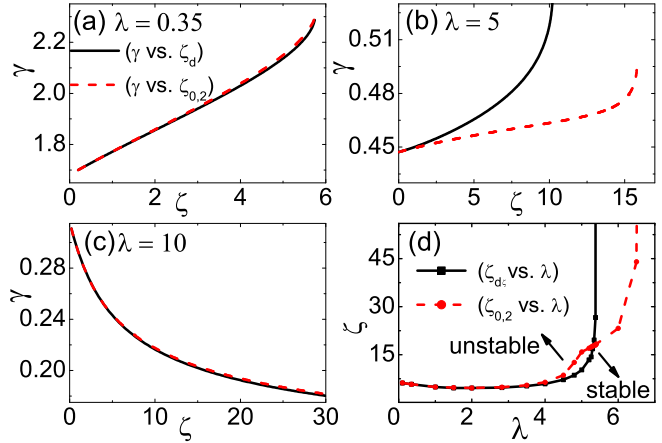}
\vspace{-0.1cm} \caption {(Color online) In frame (a)-(c), the condensate aspect
ratio, $\gamma$ (see text for the definition), is plotted for MG
against the value of dimensionless $\zeta_d$ ($\zeta_{0,2}$ set to
zero) or $\zeta_{0,2}$ ($\zeta_{d}$ set to zero) for three
trapping aspect ratio $\lambda =\omega_z/\omega_\rho=0.35,5$ and
$10$. Frame (d) shows the phase boundary separating the unstable
(collapsed) state from the stable state. In all frames,
$\zeta_{s}=0$ is taken.} \label{fig1}
\end{figure}

When the MG (\ref{eq:wave}) is used as the trial wave function, the
three terms of the variational energy become
\begin{eqnarray}
E_{\rm k} &=& \frac{{R_0^2 + 2R_z^2}}{{4R_z^2R_0^2}} \nonumber \\
&-& {B}\left( {\frac{{4{a_0}}}{{R_0^2}} + \frac{{2{a_z}}}{{R_z^2}}
-
\frac{{a_z^2}}{{R_z^2}} + 2{a_0}{a_z}\frac{{R_0^2 + R_z^2}}{{R_z^2R_0^2}}} \right),\nonumber \\
 {E_{\rm trap}} &=& \frac{1}{{{B}}}[R_0^2(2 + 8{a_0} + 2{a_z} + 12a_0^2
 + 3a_z^2 + 4{a_0}{a_z})\nonumber \\
&+& R_z^2{\lambda ^2}(1 + 2{a_0} + 3{a_z} + 2a_0^2 + 15a_z^2 + 3{a_0}{a_z})], \nonumber
 \end{eqnarray}
and
\begin{eqnarray}
{E_{{\mathop{\rm int}} }} &=& {\zeta _s}{A_s}\left( {{a_0},{a_z},{R_0},{R_z}} \right)
 + {\zeta _d}{A_d}\left( {{a_0},{a_z},{R_0},{R_z}} \right)\nonumber \\
&+& {\zeta _{0,2}}{A_{0,2}}\left( {{a_0},{a_z},{R_0},{R_z}}
\right), \label{eq:E}
 \end{eqnarray}
where $B =({4{a_z} + 8a_0^2 + 3a_z^2 + 8{a_0} + 4{a_0}{a_z} + 4})$.
$A_s$, $A_d$, and $A_{0,2}$ are lengthy functions of
${a_0},{a_z},{R_0}$, and ${R_z}$ whose explicit forms are given in
Appendix~\ref{app1}. One can simply
check that when $a_0=a_z=0$, $B\rightarrow 4$ and $E_{\rm
k}$ and $E_{\rm trap}$ reduce to those for the SG case [see
Eqs.~(\ref{eq:Ek}) and (\ref{eq:Etrap})]. $E_{\rm int}$ will
also reduce to that for the SG case (see Appendix~\ref{app1}).

As seen clearly in $E_{\rm int}$ in Eq.~(\ref{eq:E}), $\zeta_d A_d$ and
$\zeta_{0,2} A_{0,2}$ correspond to the contributions of dipolar
interaction and higher order $\Delta a_{0,2}^{(0)}$ interaction to
the ground-state energy. Thus the regime at which the two terms
become most distinct is also the regime to see the effect beyond
the FBA most clearly. In this section, we will investigate such
beyond-FBA effect from the condensate aspect ratio, $\gamma$,
which is defined as
\begin{eqnarray}
\gamma\equiv \sqrt{I_z/I_x}, \label{eq:gamma}
\end{eqnarray}
where
\begin{eqnarray}
&&I_z=\int {\left| {\Psi ({\bf{r}})} \right|^2 z^2 d{\bf{r}}}
\nonumber\\
&&I_x = \int {\left| {\Psi ({\bf{r}})} \right|^2 x^2 d{\bf{r}}}
=\int {\left| {\Psi ({\bf{r}})} \right|^2 y^2 d{\bf{r}}}.
\label{eq:I}
\end{eqnarray}

In Fig.~\ref{fig1}(a)--(c), based on the MG trial wave function,
we show results of the ground-state aspect ratio, $\gamma$, against the dipolar
interaction strength, $\zeta_d$, for $\zeta_{0,2}=0$, and against
$\zeta_{0,2}$ by setting $\zeta_{d}=0$ for three different values
of the trapping aspect ratio, $\lambda$. Note that, although such
kind of interaction is not realistic in the experiment, but it
provides a direct evidence to distinguish the effects of dipolar
interaction in the FBA and beyond the FBA.
In view of Fig.~\ref{fig1}(a) and (c), for the $\lambda=0.35$ and
$10$ cases, the effects of $\zeta_{d}$ and $\zeta_{0,2}$ are almost
the same within the parameter regime we calculate, because the
condensate aspect ratio is mostly determined by the trapping aspect
ratio directly, i.e. the interaction effect is negligible. In
contrast, for the case of $\lambda=5$ [see Fig.~\ref{fig1}(b)], the
effects of $\zeta_{d}$ and $\zeta_{0,2}$ are quite distinct: the
system becomes collapsed more easily for the case of a pure FBA
interaction, $\zeta_d$ (solid line), while it becomes less
easily collapsed for the other case (dashed line). In Fig.~\ref{fig1}(d), we
show the critical value of $\zeta$ for these two cases as a function
of the trapping aspect ratio, $\lambda$. One sees that system
becomes always stable when $\lambda$ is larger than 5.3 for the bare
FBA result, while it requires 6.5 for a pure $\Delta a_{0,2}^{(0)}$,
the leading order effect beyond the FBA. Our result indicates that
$\zeta_{0,2}$ acts more attractively than $\zeta_{d}$ in the
intermediate $\lambda$ regime.

It is useful to note that regarding where the system collapses,
there might be some quantitative difference between the variational
and numerical results. Under the condition of same $\lambda$, real
system could collapse more easily than what the variational theory predicts
in the smaller dipolar interaction regime \cite{ronen06,ronen07}. In
addition, for large $\lambda$ (the system is of pancake shape), the system
could also be unstable at large $\zeta_d$ \cite{dutta07,ronen07}.
Nevertheless, we emphasize that the qualitative features predicted by
the variational method should still be reliable for large $\zeta_d$ so long as
$\lambda$ is not too large. The latter is exactly what is studied
in the current context.

\section{Beyond the FBA: elementary excitations}\label{sec4}
\begin{figure}
\vspace{0.1cm}
\includegraphics[width=0.4\textwidth]{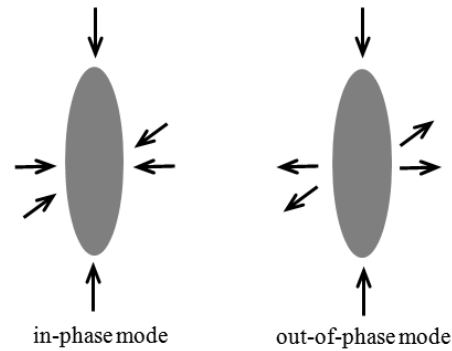}
\vspace{0.cm} \caption {Schematic plot of the in-phase and
out-of-phase breathing modes of an anisotropic dipolar gas. In
Ref.~\cite{goral02}, these two kinds of modes are called breathing
and quadrupole modes.} \label{fig2}
\end{figure}

This section devotes to the elementary excitations of a dipolar
condensate beyond the first Born approximation. As sketched in
Fig.~\ref{fig2}, two types of collective modes, namely the in-phase
and out-of-phase breathing modes, are considered here. They
are of particular importance because they are the lowest two
excitation modes in a trapped condensate \cite{baranov08,ronen06}.
In the paper of G\'{o}ral and Santos \cite{goral02}, these two modes
were also called breathing and quadrupole modes. For simplicity of
presentation, we will show the results only for the lower-energy one,
although both of them are simultaneously obtained in the same method.

In this paper, variational method is applied to study the
collective excitation of a system. It corresponds to solving the
stationary point of the action $S = \int dt L$, where the Lagrangian
$L=T-E$ with $T = \int d{\bf r}(i\hbar/2) \left[ \Psi ^*(
{\bf{r}}){\partial \Psi( {\bf{r}})}/{\partial t}-\Psi(
{\bf{r}}){\partial \Psi ^*({\bf{r}})}/{\partial t}\right]$ and $E$
is obtained by the sum of the terms in Eq.~(\ref{eq:E}). For the
study of breathing modes, the MG trial function in Eq.~(\ref{eq:wave})
will be generalized to include dynamical variables as follows:
\begin{eqnarray}
 \Psi \left( {{\bf{r}},t} \right) = C\left( t \right)
 e^{\left[ { - \frac{{\rho^2
 \left( {1 + \varepsilon_0 \left( t \right) +
 i\varepsilon '_0 \left( t \right)} \right)}}{{2R_0^2 }} -
 \frac{{z^2 \left( {1 + \varepsilon _z \left( t \right) +
 i\varepsilon '_z \left( t \right)} \right)}}{{2R_z^2 }}} \right]}\nonumber\\
\times \left( {1 + \frac{{a_0\left( {1 + \varepsilon _0 \left( t
\right)} \right)\rho^2}}{{R_0^2 }} + \frac{{a_z\left( {1 +
\varepsilon _z \left( t \right)} \right)z^2 }}{{R_z^2 }}}
 \right).\nonumber\\
 \label{eq:modified}
\end{eqnarray}
Here, in a cylindrically symmetric trap, $\varepsilon_i$ and
$\varepsilon'_i$ ($i=0,z$) correspond to the fluctuations of local
amplitude and local phase of the dipole cloud associated with the
$\rho$ and $z$ directions. As mentioned before, the values of
$R_0$, $R_z$, $a_0$, and $a_z$ are determined by minimizing the
energy functional. After some lengthy derivations, we obtain the
dispersions for the breathing modes:
\begin{eqnarray}
 {\omega ^2} &=&\{{f_1}{f_4}+{f_2}{f_5}
 \pm{[{({f_1}{f_4} - {f_2}{f_5})^2} + 4f_3^2{f_4}{f_5}]^{1/2}}\}/2,\nonumber \\
 \label{eq:breath}
\end{eqnarray}
where $\pm$ correspond to the in-phase or out-of-phase modes and
\begin{eqnarray}
 {f_1} &=& \frac{1}{2}\frac{{{\partial ^2}E}}{{\partial {R_0}^2}},~~~{f_2}
 = \frac{1}{2}\frac{{{\partial ^2}E}}{{\partial {R_z}^2}},~~~{f_3}
 = \frac{{{\partial ^2}E}}{{\partial {R_z}\partial {R_0}}},\\
 {f_4} &=& {B}{\left( {4 + 16{a_0} + 4{a_z} + 24a_0^2
 + 3a_z^2 + 8{a_0}{a_z}} \right)^{-1}}, \nonumber \\
 {f_5} &=& 2{B}{\left( {4 + 8{a_0} + 12{a_z}
 + 8a_0^2 + 15a_z^2 + 12{a_0}{a_z}} \right)^{-1}}. \nonumber
 \label{eq:breathc}
\end{eqnarray}
To determine which one corresponds to the in-phase or out-of-phase mode in
(\ref{eq:breath}) requires solving explicitly the time dependence
of the dynamical variables.

\begin{figure}
\vspace{0.0cm}
\includegraphics[width=0.45\textwidth]{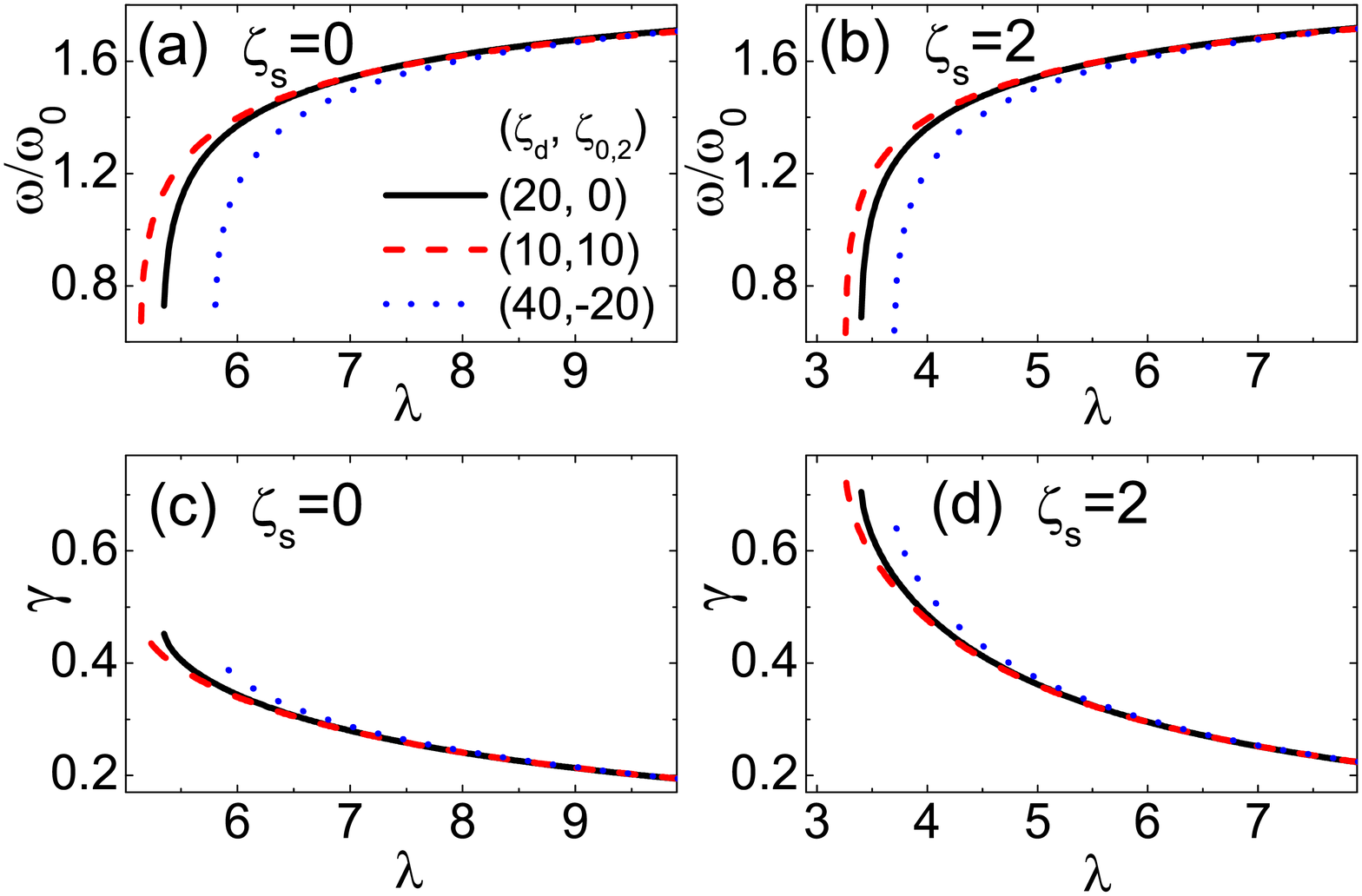}
\vspace{-0.5cm} \caption {(Color online) Frame (a) and (b): Lower breathing mode
frequencies are plotted as a function of
$\lambda$ for three combinations,
$(\zeta_{d},\zeta_{0,2})=(10,10),(20,0)$, and $(40,-20)$. Here
$\zeta_{s}=0$ for (a) and $\zeta_{s}=2$ for (b). Frame (c) and (d):
Parallel to frame (a) and (b), $\gamma$ ratios are plotted as a
function of $\lambda$.} \label{fig3}
\end{figure}

In the following, a case study is given first, which will lead to an
explicit idea how to extract quantitatively the $\zeta_{0,2}$ term
in real experiments. In $^{52}$Cr atom dipolar
BEC, it was measured that $D^2M/\hbar^2\simeq$ 24 {\AA}. If
atom number is about $N =10^4$ and the harmonic oscillator
length is about $1\mu$m, $\zeta_d$ will be about $20$. For
polar molecule systems, $\zeta_d$ could be $100$ times larger than that
of $^{52}$Cr atom dipolar BEC \cite{baranov08}. Nevertheless, here we study
the case of $\zeta_{d}+\zeta_{0,2}=20$ to which three combinations
of $(\zeta_{d},\zeta_{0,2})=(10,10),(20,0)$, and
$(40,-20)$ are considered.

In Fig.~\ref{fig3}(a) and (b), the lower
breathing modes for these three combinations are
presented and compared as a function of trapping aspect ratio
$\lambda$. In the current context, the lower breathing mode oscillates
out-of-phase. This is mainly due to the fact that $s$-wave interaction
is relatively small and the dipolar interaction dominates.
Since dipolar interaction
is anisotropic: attractive along the $z$ direction while repulsive in
the $xy$ plane, consequently out-of-phase mode has lower
energy than that of the in-phase mode.
Fig.~\ref{fig3}(a) corresponds to the case of
$\zeta_{s}=0$. In view of Fig.~\ref{fig3}(a), two important features
are revealed: (i) all three curves merge in the large $\lambda$
limit (i.e. a pancake like trapping potential) and (ii) the curves
deviate each other significantly when the system is close to the
phase boundary of collapse. For instances
when $\lambda=5.9$,  $\omega=1.338\omega_0$ for the case
$(\zeta_{d},\zeta_{0,2})=(20,0)$, while $\omega=1.044\omega_0$ for the
case $(\zeta_{d},\zeta_{0,2})=(40,-20)$. The relative frequency
difference, $\Delta\omega\equiv |\omega_1-\omega_2|/\omega_2$, can
be $21\%$ large. $\Delta\omega$ will increase even more
significantly when the system is approaching the collapsed
regime. Similar behaviors are also found in Fig.~\ref{fig3}(b),
where a finite value of $s$-wave scattering length is included
($\zeta_s=2$).

In Fig.~\ref{fig3}(c) and (d), we show the condensate aspect
ratio, $\gamma$, as a function of $\lambda$. In view of
Fig.~\ref{fig3}(c) with $\zeta_{s}=0$, one finds that all three
$\gamma$ curves also merge in the large $\lambda$ limit. Besides,
when $\lambda=5.9$ close to the collapsed regime, $\gamma=0.353$
and $0.392$ respectively for the case of
$(\zeta_{d},\zeta_{0,2})=(20,0)$ and $(40,-20)$. The relative
$\gamma$ difference, $\Delta\gamma\equiv
|\gamma_1-\gamma_2|/\gamma_2$, is about $11\%$. In
Fig.~\ref{fig3}(d) with $\zeta_{s}=2$, when $\lambda=3.8$,
$\gamma$ ratio is $0.528$ and $0.593$ for the case of
$(\zeta_{d},\zeta_{0,2})=(20,0)$ and $(40,-20)$. This gives a
relatively smaller $\Delta\gamma=2\%$. Similar to the
$\Delta\omega$ case, $\Delta\gamma$ will increase significantly
when the system is even more close to the collapsed regime.

It should be emphasized that in the case that $\zeta_{s}$ is large
or the sum of $\zeta_{d}+\zeta_{0,2}$ is small, the system will
tend to stabilize over a large span of $\lambda$. This means that
whatever combinations of $\zeta_{d}$ and $\zeta_{0,2}$ under
$\zeta_{d}+\zeta_{0,2}={\rm const}$ will roughly lead to the same
curve. Thus, $\Delta\omega$ or $\Delta\gamma$ studied above, will
always be small, i.e. the effect of the effects beyond the FBA cannot
be observed easily.

In the following, we propose how to extract quantitatively
the value of $\zeta_{0,2}$ by measuring the lower breathing mode
frequency or the condensate aspect ratio $\gamma$ of the system.
Firstly, we assume that the value of $\zeta_{s}$ is relatively
small such that the behaviors of breathing mode frequency and
$\gamma$ ratio resemble those in Fig.~\ref{fig3}(a)-(d). Suppose
that one does not know the value of any one of $\zeta_{s}$,
$\zeta_{d}$, and $\zeta_{0,2}$. One can first perform the
measurement of the lower breathing mode frequency and/or $\gamma$
ratio at large $\lambda$ case, say $\lambda=10$. As shown in
Fig.~\ref{fig3}(a)-(d), all three curves merge at large $\lambda$
limit. This means that one can unambiguously determine the values
of both $\zeta_{s}$ and the sum of $\zeta_{d}$ and $\zeta_{0,2}$
simply by carrying out a theoretical fitting to the experimental
data. The next task is to separate the values of $\zeta_{d}$ and
$\zeta_{0,2}$. For this purpose, one can redo the experiment and
decrease $\lambda$ to approach the collapsed regime. Since at this
regime, the fitting will be quite sensitive to both values of
$\zeta_{d}$ and $\zeta_{0,2}$,  one can then unambiguously
determine the value of $\zeta_{0,2}$ by comparing with our
theoretical calculation result.

\begin{figure}[tbp]
\vspace{0.0cm}
\includegraphics[width=0.45\textwidth]{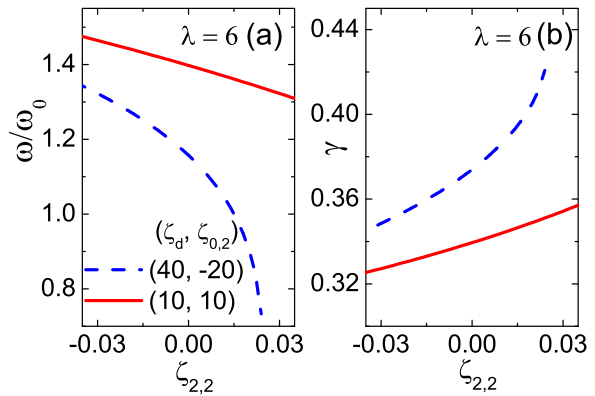}
\vspace{0.0cm} \caption {(Color online) (a) Lower breathing mode
and (b) $\gamma$ ratio of the dipolar system are plotted as a
function of $\zeta_{2,2}$. In both frames, $\lambda=6$ and two
combinations of $(\zeta_{d},\zeta_{0,2})=(10,10)$ and $(40,-20)$ are
chosen. Here we set $\zeta_s=0$.} \label{fig4}
\end{figure}

Now we provide a short discussion on the effect of the next order
effect beyond the FBA by including finite value of
$\Delta{a}_{2,2}^{(0)}$. Same as before, we define
$\zeta_{2,2}\equiv N\Delta{a}_{2,2}^{(0)}$ for the convenience. In
Fig.~\ref{fig4} we show the lower breathing mode frequency and the aspect
ratio, $\gamma$, as a function of $\zeta_{2,2}$ for a fixed trapping
aspect ratio, $\lambda=6$. Two combinations of
$(\zeta_{d},\zeta_{0,2})=(10,10)$ and $(40,-20)$ are chosen and it is assumed
that $\zeta_s=0$. As $\zeta_{2,2}$ increases, it is found that the system
becomes more cigar-like shape (i.e. $\gamma$ becomes larger), while
it becomes more pancake-like shape when $\zeta_{2,2}$ is negative.
Positive $\zeta_{2,2}$ will lead the system to be more
attractive along the $z$ direction and hence more easily collapsed. However its
effect is relatively weaker compared to that of
the $\zeta_{0,2}$ term discussed before.

\section{Conclusion} \label{sec5}

This paper attempts to study the ground state and elementary
excitations of a strongly interacting dipolar bosonic gas based on a
theory going beyond the first Born approximation (FBA). By using an
appropriate trial wave function in the variational method, the
leading higher-order corrections to the FBA are studied in details,
in particular for the condensate aspect ratio and the elementary
excitation mode frequency. Of most interest, it is found that the effect of the
corrections is most important when the system is close to phase
boundary of collapse. Through several case studies with parameters
pertaining to $^{52}$Cr atom dipolar BEC, it is believed that the
higher-order effect going beyond FBA should be even more significant
and highly observable in the strong dipolar molecule system.
An idea for extracting quantitatively such leading-order effect
beyond the FBA in real experiments is provided. Finally, to
shed more light on understanding the higher-order
effect, a more accurate numerical calculation is
in demand.

\acknowledgements

We are grateful to the support of National Science Council and
National Center for Theoretical Sciences, Taiwan.

\appendix
\section{Forms of Functions A}\label{app1}
In this Appendix, we give explicit forms of functions $A_s$, $A_d$, and $A_{0,2}$
appearing in Eq.~(\ref{eq:E}).
Function $A_s$ is analytic and given by
\begin{eqnarray}
A_s &=& \frac{{\sqrt 2 }}{{\sqrt \pi  {R_z}R_0^2{{\left( {8\,a_0^2 + 4\,{a_z}{a_0} + 8a0 + 4\,az + 3\,a{z^2} + 4} \right)}^2}}} \nonumber\\
 &\times& \left[ {12a_0^4 + \left( {24 + 6\,{a_z}} \right)a_0^3 + \left( {24 + 12\,{a_z} + \frac{9}{2}\,a_z^2} \right)a_0^2 } \right. \nonumber\\
 &+& \left( {16 + 12\,{a_z} + 9a_z^2 + \frac{{15}}{4}a_z^3}
 \right){a_0} \nonumber \\
 &+& \left. {  \frac{{105\,}}{{32}}a_z^4 + \frac{{15}}{2}a_z^3 + 9\,a_z^2 + 8\,{a_z}+8} \right].
 \label{eq:apendzets}
\end{eqnarray}
Functions $A_d$ and $A_{0,2}$ are represented by an integral in the
cylindrical coordinate, $(k_r, k_\varphi, k_z)$. More
precisely $A_d$ is given by
\begin{eqnarray}
A_d=\int\limits_{-\infty} ^{ \infty }d{k_z}
{\int\limits_0^\infty{2\pi {k_r}d{k_r}} ~{\eta _d}}
\end{eqnarray}
with
\begin{eqnarray}
\eta_d&=& \frac{{{e^{ - (R_z^2k_z^2 + R_0^2k_r^2)/2\,}{(3k_z^2/(k_r^2 + k_z^2) - 1)}}}}{{12\pi^2(4\,{a_0}\,{a_z} + 8\,{a_0} + 8\,a_0^2 + 4 + 4\,{a_z} + 3\,a_z^2)^2}}\nonumber  \\
  &\times& \left[ {4\,{a_0}\,{a_z} + 8\,{a_0} + 8\,a_0^2 + 4 + 4\,{a_z} + 3\,a_z^2} \right.\nonumber \\
  &-&\left( {2\,{a_0} + 2 + 3\,{a_z}} \right){a_z}R_z^2k_z^2- \left( {2 + 4\,{a_0} + {a_z}} \right){a_0}R_0^2k_r^2 \nonumber \\
 &+&\left. {  \frac{{a_z^2R_z^4k_z^4}}{4} + \frac{{a_0^2R_0^4k_r^4}}{4} + \,\frac{{{a_0}\,{a_z}\,R_0^2R_z^2k_z^2k_r^2}}{2}}
 \right]^2,
 \label{eq:apendzetad}
\end{eqnarray}
while $A_{0,2}$ is given by
\begin{eqnarray}
A_{0,2}= \int\limits_{-\infty} ^{ \infty }d{k_z}
{\int\limits_0^\infty{2\pi {k_r}d{k_r}}~{\eta_{0,2}} }
\end{eqnarray}
with
\begin{eqnarray}
\eta_{0,2}&=& \frac{{{e^{ - (R_z^2k_z^2 - R_0^2k_r^2)/2}}(3k_z^2/(k_r^2 + k_z^2) - 1)}}{{12{\pi ^2}{{\left( {4\,{a_0}\,{a_z} + 8\,{a_0} + 8\,a_0^2 + 4 + 4\,{a_z} + 3\,a_z^2} \right)}^2}}}\nonumber \\
  &\times& \left[{ {C_{rz}}{a_0}{a_z}R_0^2R_z^2k_r^2k_z^2+ {D_r}a_0^2R_0^4k_r^4 + {D_z}a_z^2R_z^4k_z^4 }\right. \nonumber\\
  &+&\left. { {C_0} + {C_r}{a_0}R_0^2k_r^2 + {C_z}{a_z}R_z^2k_z^2}
  \right].
 \label{eq:apendzeta02}
\end{eqnarray}
Here $C_0$, $C_r$, $C_z$, $C_{rz}$, $D_r$, and $D_z$ are given as follows:
\begin{eqnarray}
 C_0&=&40\,a_0^4 + \left( {44\,{a_z} + 96} \right)a_0^3 + \left( {33\,a_z^2 + 96\,{a_z} + 112} \right)a_0^2\nonumber \\
  &+& \left( {\frac{{39}}{2}\,a_z^3 + 60\,a_z^2 + 88\,{a_z} + 64} \right){a_0}\nonumber \\
  &+& \frac{{81}}{{16}}\,a_z^4 + 18\,a_z^3 + 34\,a_z^2 + 32\,{a_z} +
  16, \nonumber\\
 C_r= &-& \left( {\frac{{21\,}}{4}a_z^3 + 11\,{a_0}\,a_z^2 + 13\,a_z^2 + 14\,a_0^2{a_z} + 32\,{a_0}\,{a_z}} \right. \nonumber\\
 &+& \left. { 20\,{a_z} + 40\,a_0^2 + 48\,{a_0} + 16 + 16\,a_0^3}
 \right),\nonumber \\
  C_z= &-& \left( {\frac{3}{4}\,a_z^3 + 9\,a_z^2 + \frac{{15}}{2}\,{a_0}\,a_z^2 + 18\,a_0^2{a_z} + 28\,{a_z}} \right.\nonumber \\
 &+& \left. { 36{a_{0\,}}{a_z} + 20\,a_0^3 + 40\,a_0^2 + 40\,{a_0} + 16}
 \right), \nonumber\\
 C_{r,z}&=&\left( {\frac{3}{2}\,a_z^2 + 4\,{a_z} + 2\,{a_0}\,{a_z} + 8 + 8\,{a_0} + 4\,a_0^2}
 \right),\nonumber\\
 D_{r} &=& \left( {\frac{3}{4}\,a_z^2 + 2{a_z} + \,{a_0}\,{a_z} + 4 + 4\,{a_0} + 2\,a_0^2} \right),\nonumber \\
 D_{z} &=& \left( {\frac{{3\,}}{4}a_z^2 + 2\,{a_z} + \,{a_0}\,{a_z} + 4 + 4\,{a_0} + 2\,a_0^2}
 \right).
\label{eq:apendzeta021}
\end{eqnarray}
When $a_0=a_z=0$, the two integrals for $A_d$ and $A_{0,2}$ become equal.

\end{document}